# Unlocking Adaptive User Experience with Generative AI

Yutan Huang[a], Tanjila Kanij[b], Anuradha Madugalla[c], Shruti Mahajan, Chetan Arora[d], and John Grundy[e]

*Department of Software Systems and Cybersecurity, Monash University, Clayton, Melbourne, Australia*
{*yutan.huang1,tanjila.kanij,anu.madugalla*}@*monash.edu,shrutimahajan271@gmail.com,*
{*chetan.arora,john.grundy*}@*monash.edu*

Abstract: Developing user-centred applications that address diverse user needs requires rigorous user research. This is time, effort and cost-consuming. With the recent rise of generative AI techniques based on Large Language Models (LLMs), there is a possibility that these powerful tools can be used to develop adaptive interfaces. This paper presents a novel approach to develop user personas and adaptive interface candidates for a specific domain using ChatGPT. We develop user personas and adaptive interfaces using both ChatGPT and a traditional manual process and compare these outcomes. To obtain data for the personas we collected data from 37 survey participants and 4 interviews in collaboration with a not-for-profit organisation. The comparison of ChatGPT generated content and manual content indicates promising results that encourage using LLMs in the adaptive interfaces design process.

## 1 INTRODUCTION

Designing and developing user-friendly web interfaces is both an art and science. User Interface (UI) and User Experience (UX) design requires striking a balance between aesthetic appeal and meeting functional and non-functional requirements. The requirements will be based on a deep understanding of user behaviour, expectations and special 'human-centric' needs. These needs are often diverse and individualised (Benyon, 2019). The traditional design processes, while structured and methodical, are characterised by their resource-intensive nature. They often require input from multidisciplinary teams (domain experts, software engineers and UI/UX designers) and provides solutions that may not effectively address the individual requirements of diverse users (Lewis and Sauro, 2021). In addition, there is a growing need for swift time-to-market of web applications (web apps), which has led to a pressing demand for innovative, efficient and more adaptable UI/UX design methodologies in the industry (Main and Grierson, 2020).

The recent advancements in artificial intelligence (AI) techniques, particularly generative AI and large language models (LLMs) like ChatGPT, mark the beginning of potential new innovations in several areas, including UI/UX designs. These technologies offer a promising shift toward more personalised and adaptable design processes, paving the way for innovative solutions to longstanding issues. (Wang et al., 2023; Nguyen-Duc et al., 2023).

In this new ideas paper we introduce a new approach to develop user personas and UI/UX for web applications by leveraging the generative capabilities of ChatGPT. We explore generating an adaptive UI with LLMs by providing user personas as an input to the design process. We then compare these results with another set of adaptive UIs that were manually developed based on rigorous user research. We compare ChatGPT-generated outcomes with those crafted through conventional user research and design practices. We show that the trialled approach offers a possibility of a 'true' human-centred design of web apps that could streamline design processes and tailor user experiences more closely to individual needs. The results reflects an anticipating stance, inviting further research and discussion from the community and aims to set stage for future empirical research to extend and validate our preliminary findings. This paper is organised as follows: Section 2 reviews existing literature, Section 3 outlines our research design and methodol-ogy, Section 4 describes our preliminary results and presents a comparative analysis of the results. Section 5 presents a SWOT analysis, reflecting on the strengths, weaknesses, opportunities, and threats of using LLMs in the design process based on our outcomes. Section 7 presents our conclusions and future research direction ideas for using generative AI in adaptive UI/UX design.

---

[a] https://orcid.org/0000-0002-6239-9665
[b] https://orcid.org/0000-0002-5293-1718
[c] https://orcid.org/0000-0002-3813-8254
[d] https://orcid.org/0000-0003-1466-7386
[e] https://orcid.org/0000-0003-4928-7076

## 2 RELATED WORK

### 2.1 Generative AI

**ChatGPT:** Generative Pre-trained Transformer (GPT) is an autoregressive large language model that produces human-like text through deep learning (Atlas, 2023). The latest version released in 2023, GPT 3.5 has been shown to achieve a significant improvement compared to its previous GPT models (Talebi et al., 2023). ChatGPT platform developed by OpenAI uses GPT 3.5 and is considered one of the best LLMs (Magar and Schwartz, 2022). As such we selected ChatGPT 3.5 as the LLM for this study.

**Prompt Engineering with ChatGPT:** Prompt engineering refers to the set of instructions provided to LLMs to receive responses and serves as an essential process in leveraging ChatGPT to generate meaningful and contextually relevant outputs for UI/UX design and other SE activities (Liu et al., 2023). The OpenAI Playground, a popular tool for experimenting with prompts and examples, provides significant value to users. However, there is currently little guidance on systematically crafting prompts in a rigorous manner (Arora et al., 2023). Consequently, we have incorporated prompting techniques based on rigorous evaluation by the researchers, our methodology involved a deliberate and iterative approach to crafting prompts that would lead to the persona generation and web page designs that aligns closely with user needs. For prompt engineering with ChatGPT, we used the output customisation and error identification category from the catalogue suggested by White et al. (White et al., 2023a). The crafting of prompts for ChatGPT ensures creating personas and web pages that are comparable, if not superior, to those developed through traditional methods.

### 2.2 Persona

A persona is a fictional character that reflects the characteristics of clusters of real end users, they are often utilised for analysing end-user requirements and ensuring that the proposed software products ultimately meet these standards (Karolita et al., 2023). There are no universally accepted methods for creating personas, and researchers have proposed various approaches to develop personas for user groups. These approaches fall into three categories: qualitative, quantitative, and mixed (Tu et al., 2010). Qualitative methods for persona creation depend on exploratory research with a medium-sized sample of users. This method involves users at all stages of design, focuses on understanding and analysing user behavior, and iterates the creation of personas (Hosono et al., 2009). On the other hand, quantitative methods aim to leverage user data from diverse sources to construct personas, thereby improving understanding of users. Mixed methods combine elements from both quantitative and qualitative approaches, often targeting specific user demographics, such as groups defined by age (Cooper et al., 2007).

### 2.3 Nielsen's Persona Method

We employ a generalised mixed method for creating personas, with Nielsen's ten-step approach as our manual method (Nielsen, 2004). We selected Nielsen's approach as our information does not align perfectly with either the qualitative or quantitative approaches. By following Nielsen's structured process, we ensure that our personas are well-founded, accurately reflecting the complex needs, attitudes, and behaviors of our target user groups.

Nielsen's ten-step approach to creating personas encompasses three crucial domains: data collection, engagement with persona descriptions, and organisational buy-in. Initially, it involves gathering data from various sources to understand target users, including the methods for data collection and storage. After collecting data, an initial hypothesis about the users is formed to guide the persona creation process. This hypothesis is then tested by verifying the data. As patterns within user groups emerge, these groups are categorised, leading to the construction of personas equipped with detailed backgrounds and personal traits. It's essential to place these personas in specific scenarios and validate them with real-life users to start creating scenarios that are relevant to the personas. Once validated, these personas are shared with organisations to assist developers. The final step involves crafting a narrative for the persona, detailing scenarios that highlight the persona's goals.

## 3 RESEARCH DESIGN

### 3.1 Methodology Overview

We want to answer the following key research questions (RQs):

1. **RQ1 -** How do personas developed by LLMs perform against the manually created personas?
2. **RQ2 -** What is the quality of web pages developed with the use of LLMs?
3. **RQ3 -** What is the effectiveness of the customized UI/UX in web pages developed by LLMs based on specific personas?

To address these RQs, we adopted a mixed-method research design, integrating quantitative data from surveys with qualitative insights from interviews. This blended approach facilitated a comprehensive comparison between the innovative use of ChatGPT in generating personas and web pages, and the traditional manual methodologies employed in UI/UX design. The research was designed in a phased approach, with distinct stages aimed at directly addressing each research question: **Persona development phase -** This examines the depth, accuracy and applicability of ChatGPT-generated personas in comparison to those crafted manually, which addresses RQ1. **Webpage design phase -** This addresses RQ2 by evaluating the quality of ChatGPT-developed web pages with user representatives, focusing on their design principles, aesthetic appeal, as well as user engagement metrics. **UI/UX customization phase -** This addresses RQ3 by exploring the effectiveness of ChatGPT in tailoring web pages to specific user personas, assessing ChatGPT's adaptability and precision in meeting user-specific design requirements.

## 3.2 Survey and Interview

We partnered with an NFP organisation and conducted a survey (with 37 participants) and interviews (with four participants) to obtain the user preferences and needs for their website. The NFP is a community support organisation providing health and wellness support to users, including organising events and workshops that they advertise through the website (in question) and wanted people to register through the website as well. Participants were recruited by convenience sampling with the help of the partnered organisation. The survey setup with Google Forms, collected data on demographic information, the daily usage of websites and web applications, and their feedback on the website of the partner organisation. Interviews were conducted through the Zoom platform where participants were asked to explore the organisation's website and provide suggestions to improve the website design. They were then instructed to write the features and information they desired on empty stickers on the Miro board, an online collaborative whiteboard platform. The insights gathered inform the criteria for evaluating the effectiveness of ChatGPT-generated outputs against manual efforts, they were instrumental in guiding the development of user personas, ensuring the outputs of our study to be grounded in real-world environments.

## 4 RESULTS

### 4.1 RQ1 Results

#### 4.1.1 Persona development with ChatGPT

For persona development, we implemented a comprehensive prompt engineering strategy that involved the integration of user scenarios and characteristics derived from preliminary user research (Zhang et al., 2023). Our design process recognises the importance of specificity in prompts to obtain high-quality outputs; we utilised the "output customisation and error identification" pattern as suggested by White et al. (White et al., 2023a) to refine our prompts based on initial feedback loops with our model. An example prompt for persona creation was structured as follows: *Given a not-for-profit organisation aiming to increase engagement among its diverse user base, create a persona representing a typical user. Consider age, interests, digital literacy, and potential engagement barriers. The persona should reflect the organisation's focus on health and wellness support, including attending workshops and events.* This prompt was designed to encapsulate the diverse user base and the organisation's mission, ensuring the personas generated were diverse and represented real-world users. We then used iterative refinement to adjust the level of detail and scope of the personas, ensuring they served as foundations for further design decisions. A persona generated by ChatGPT using our prompt is given on the left-hand side in Figure 1.

#### 4.1.2 Persona development from survey

We applied Nielsen's approach for persona development from the survey data (Nielsen, 2004). Following the approach, we found three main clusters of user facets from the survey results. **Cluster 1:** Consists of end-users from the age group 56-65, who are mostly retired individuals with various cultural backgrounds. They experience difficulties navigating websites and prefer bigger texts and simple layouts; they browse websites for news media and register for community activities on the organisation's website. **Cluster 2:** Comprises users aged 26-35; most come from Chinese cultural backgrounds and various occupations. They experience little difficulty navigating websites and, like content-rich websites with a range of vibrant colours, they often register for workshops and social activities on the organisation's website. **Cluster 3:** Consists of end-users from the age group of 18-26, who are primarily students from Asian cultural backgrounds. They browse websites with no difficulties and prefer a monochromatic colour for websites; they register and attend stress management workshops specifically on the organisation's website. Three personas were developed from each cluster; named Jack, Linda and James. Linda persona is presented on the right-hand side in Figure 1.

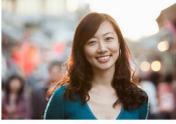

Figure 1: Persona examples (left) generated by ChatGPT and (right) by following Nielsen's approach manually.

### 4.1.3 Comparative Analysis

We summarise our comparison of manually developed personas vs ChatGPT generated personas in Table 1. The demographics displayed significant similarities, albeit with variations in titles. The manually-created personas included visual avatars and notes on cultural distinctions, enhancing their relatability. In contrast, the ChatGPT personas offered an in-depth exploration of the goals and needs specific to each persona, attached with a dedicated section titles 'How Organisation X's Website Helps,' which mapped the website's functions to the persona's daily life.

The manual approach reflected a broad classification suitable for larger user groups and the ChatGPT personas suggested a more personalised UI/UX design approach. This suggests that generative AI can complement user research methods by providing a deeper, data-driven understanding of user segments with highlighted unique user preferences and engagement barriers. It also highlights the need for a combination of AI efficiency with human-centric insights. We aim to explore this new research area more in our future studies.

Table 1: Persona features developed manually and with ChatGPT (*MP - Manual Persona, *CP- ChatGPT Persona)

| Category | Feature | MP* | CP* |
|---|---|---|---|
| Demographics | Age | ✓ | ✓ |
| | Gender | ✓ | ✓ |
| | Cultural Affiliation | ✓ | ✗ |
| | Work | ✓ | ✓ |
| | Location | ✓ | ✓ |
| | Family | ✓ | ✓ |
| | Background | ✓ | ✓ |
| Motivations | Goals | ✓ | ✓ |
| | Challenges | ✗ | ✓ |
| | Frustrations | ✓ | ✗ |
| User Needs and Preferences | Website preferences | ✓ | ✗ |
| | Colour preferences | ✓ | ✗ |
| | User Environment | ✓ | ✓ |
| | Organisation X's role | ✗ | ✓ |

## 4.2 RQ2 Results

### 4.2.1 Webpage development with ChatGPT

To develop webpages for our partner NFP organisation, we followed a two-step prompt engineering process. In the first step, we furnished ChatGPT with essential background information regarding the organisation, website structure, and details about each webpage. For example, we provided the details of the purpose and functionality of the NFP organisation's website, how many web pages they have, and the description and purpose of each web page, structuring prompts that described the desired features and content of each page clearly. In the second step, we provided a clear and specific prompt for the desired output, i.e., the generation of HTML scripts and CSS features for the website designs. In these prompts we provided the following information to ChatGPT. **Background Information** -"Create a website for a not-for-profit organisation that provides workshops and events to enrich community health and wellness. The website aims to provide support and care for the community along with in-person services." **Web-** site Information - "The website should have six web pages: A home page, an About page, a Courses page, a Services page, a Resources page, and a Donate page with a consistent colour scheme." **Home Page** - "For the home page: A "Menu" option should be available on either the top or the side of the page; a consistent color scheme should be chosen for the website; a hero feature, which is a unique feature to the home page, should be included."

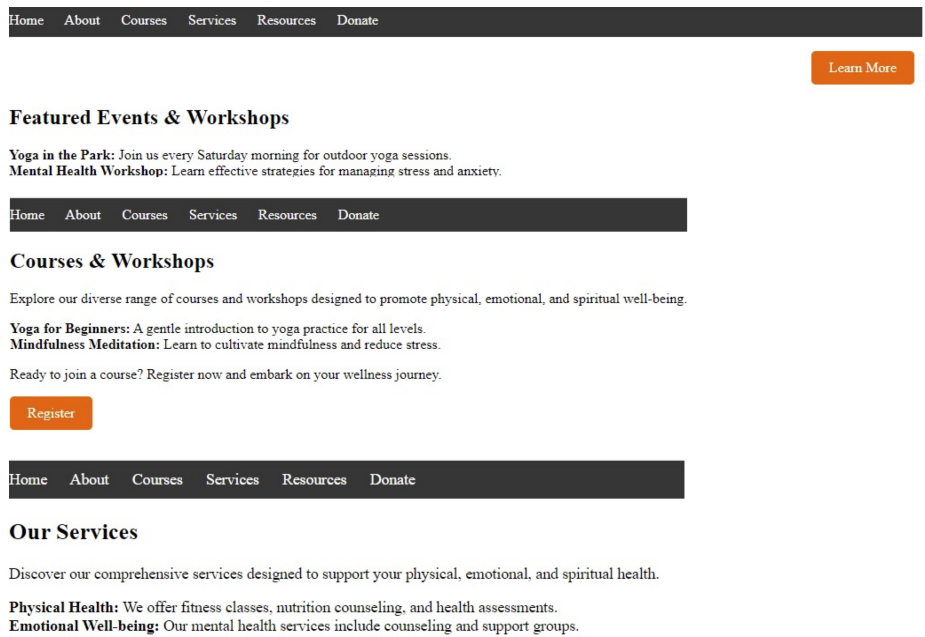

Figure 2: Webpage created by ChatGPT

We provided detailed specifications in addition to the prompt template to complete the webpage generation process. We were able to guide ChatGPT towards generating webpage designs that were not only appealing but also functional and aligned with the organisation's goals by iterating on the prompts based on initial outputs. A snapshot of the homepage generated by ChatGPT is presented in Figure 2.

### 4.2.2 Website development from interviews

Participants were requested to visit organisation's website to make the interview questions consistent with the prompt engineering for ChatGPT. The interview questions were crafted based on the analysis of survey responses. The participants were inquired about their preferences regarding the website's landing page, its level of dynamism, customisability, and how information is represented. Based on their preference, the review of the website's content and UI/UX, they were asked to share their design insights on Miroboard. One example design created on the MiroBoard is given in Figure 3.

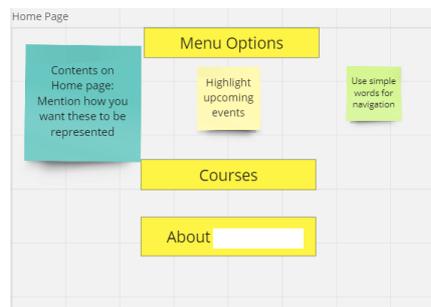

Figure 3: Miroboard webpage design

### 4.2.3 Comparative Analysis

We compared key differences and similarities between web pages manually designed by our interviewees, visualised through Miroboard, and those ChatGPT generated. One significant observation is Chat-GPT's consistent inclusion of explicit title specifications, a detail frequently overlooked by out interviewees. This aspect highlights ChatGPT's capability to maintain essential web design elements from prompts even when they are not directly mentioned, which adheres to structured content creation.

Our analysis revealed limitations in ChatGPT's ability to meet specific user expectations, notably the absence of anticipated elements e.g a video on 'about page' and contact details on the 'donation page'. This show a challenge in incorporating implicit user pref-erences. Additionally, aligning ChatGPT's designs with detailed user requirements, like clear navigation labels, reduced scrolling, and informational pop-ups, proved difficult. However, ChatGPT's designs demonstrated greater structure and consistency, benefiting from user-centered design principles. Features like interactive buttons that change color upon hovering enhanced the user experience.

Table 2: Features Comparison

| Categories | Website Features | Website developed by ChatGPT | Miroboard Design 1 | Miroboard Design 2 | Miroboard Design 3 |
|---|---|---|---|---|---|
| Home Page | Title | ✓ | ✓ | ✗ | ✓ |
|  | Menu options | ✓ | ✓ | ✓ | ✓ |
|  | Content Highlights | ✗ | ✓ | ✓ | ✓ |
|  | Courses & Events description | ✓ | ✗ | ✗ | ✓ |
|  | Simple Words for navigation | ✗ | ✓ | ✗ | ✓ |
| About Page | Title | ✓ | ✗ | ✗ | ✓ |
|  | Vision | ✓ | ✓ | ✗ | ✗ |
|  | Motivation | ✓ | ✓ | ✗ | ✓ |
|  | Videos | ✗ | ✓ | ✓ | ✓ |
| Courses Page | Title | ✓ | ✓ | ✗ | ✓ |
|  | Courses & Events details | ✓ | ✗ | ✓ | ✓ |
|  | Registration | ✓ | ✓ | ✓ | ✓ |
|  | Less Screen Scrolling | ✗ | ✓ | ✓ | ✗ |
|  | Sub menu for course separations | ✗ | ✓ | ✓ | ✓ |
| Resources Page | Title | ✓ | ✗ | ✗ | ✗ |
|  | Resource categorization | ✗ | ✓ | ✓ | ✓ |
|  | Details of resources | ✓ | ✗ | ✓ | ✓ |
| Services Page | Title | ✓ | ✓ | ✗ | ✗ |
|  | Pop-ups for expert information | ✗ | ✓ | ✗ | ✗ |
|  | More appealing titles | ✗ | ✓ | ✗ | ✓ |
|  | Service categorization | ✗ | ✓ | ✓ | ✓ |
|  | Details of services | ✓ | ✗ | ✗ | ✓ |
| Donate Page | Title | ✓ | ✗ | ✗ | ✗ |
|  | Contact details | ✗ | ✓ | ✓ | ✓ |
|  | Donation Methods | ✓ | ✗ | ✓ | ✓ |

### 4.3 RQ3 Results

In addressing RQ3, we provided ChatGPT with the three personas we developed manually based on our survey. This step aimed to understand how well ChatGPT can customise the UI/UX based on the user groups described by real-life personas. We provided one persona at a time and found ChatGPT developed web pages with the same content but differing colour themes and styles. Figure 4 presents a snapshot of three web pages ChatGPT developed for Jack, Linda and James, respectively. Web pages for Linda and James used colour themes consistent with their colour preferences. However, Jack's persona had hyperlinks instead of interactive buttons with no colour theme. This can be because Jack belongs to an elderly adult group, and web pages a few decades ago were mainly designed with hyperlinks.

Furthermore, we gathered all manually developed personas and asked ChatGPT to develop a webpage for them as a whole. This resulted in ChatGPT keeping the structural design features like interactive buttons and menu options shown in Figure 4. However, it integrated the colour preferences of all personas and developed a light green-yellow colour for interactive buttons and menu options, with the text colour changing to blue when hovering.

These results indicate that ChatGPT is capable of cuztomizing web pages to to reflect distinct user profiles, adapting to design elements like color themes and navigation features, which highlights the significance of LLMs in enabling user-sentric design approaches.

## 5 DISCUSSION

LLM based Generative AI tools such as ChatGPT, offer features that can be used in many development tasks, such as developing adaptive UI/UX. However, these opportunities are yet to be explored in detail. Bartao and Joo found that UI/UX developers do not widely use AI tools (Bertão and Joo, 2021). The recent research which incorporated ChatGPT in prototype designing found some benefits of the approach; but received mixed responses from developers (Ekvall and Winnberg, 2023). Another research with an older version of GPT found that incorporating LLMs in earlier prototyping stages can save effort and cost (Bilgram and Laarmann, 2023). However, none of the research systematically compared ChatGPT's outcome at each step based on specific prompt engineering with a traditional manual process. Our preliminary attempt to do a comparative analysis shows several promising opportunities. The persona developed by ChatGPT based on domain specification (in this case - the NFP organisation) was detailed enough and even contained a section exploring websites integration to Persona user's life, which the manually developed persona didn't have. The differences in cultural background and other preferences presented with three manually developed personas were absent in one persona developed by ChatGPT. We think this can be alleviated with specific instructions to develop more than one persona.

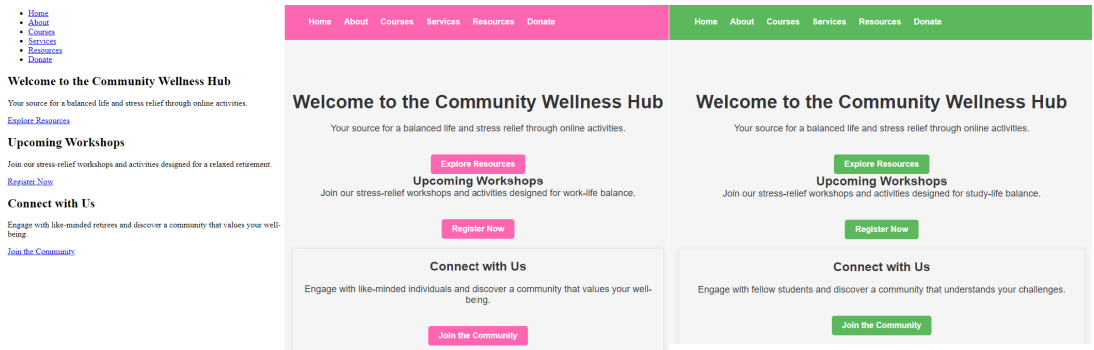

Figure 4: Webpages developed by ChatGPT with Personas

Regarding the webpages, ChatGPT developed consistent designs with different colour themes for different personas. The style was also adapted to classical style for the elderly persona. These findings indicate the opportunities ChatGPT presents to customize UI/UX based on specific user needs and preferences. Prompts can be specifically tailored to achieve these outcomes. However, tailoring prompts can be challenging and the heavy dependence of the outcome on prompt engineering poses a threat to the approach.

Our initial findings indicate promising results for using ChatGPT to develop user persona and UI/UX customized towards the persona. This can help avoid extensive user research to understand user needs and preferences and to develop UI/UX customized for the users. Based on the findings, we developed a SWOT matrix for developing UI/UX based on customized user needs and preferences with the help of ChatGPT, shown in Figure 5.

Prompt engineering is critical for ChatGPT, as the quality of prompts directly affects the relevance and quality of responses generated by ChatGPT (Yu et al., 2023). This is particularly evident in the persona development process, where we found that minor deviations in prompt choices can lead to substantial deviations in persona contexts (White et al., 2023b; Ubani et al., 2023). We also learned that prompts need to be provided within specific structural frames to achieve desired responses, without which ChatGPT would only produce the simplest forms of responses. Therefore, rigorous prompts are necessary for developing personas and web pages using ChatGPT.

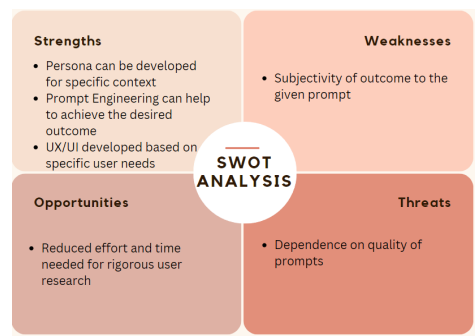

Figure 5: SWOT analysis of ChatGPT

## 6 THREATS TO VALIDITY

A potential threat to our findings are in the evaluation method as there may be a possibility of subjective bias in qualitative assessments within our study. To mitigate this threat, we plan to refine our evaluation process by involving domain experts and user representatives. Additionally, we will employ a blind review process among the evaluators to ensure they are unaware of whether the webpages and personas were generated by ChatGPT or developed manually.

Another threat is that our results may be limited by the specific context of our study—a health and wellness NFP organization. It is necessary for us to conduct further research across diverse domains to validate the broad applicability of our findings.

## 7 CONCLUSION AND FUTURE PLANS

We investigated whether ChatGPT can develop adaptive UI/UX. From our experiments, persona and website development with LLMs can be more efficient with tailored prompts being used. LLMs can generate desired outputs for developers in a short time, also providing more details and insights for outputs. The traditional approach of using quantitative and qualitative user study is time consuming, but effective for developing lightweight personas and websites. We plan to broaden our research to encompass additional aspects of generative AI in UI/UX design. We will work towards creating a robust framework to automate persona and website development, aiming to capture critical user-centric content that aids designers. Such a framework will not only streamline the design process but also serve to fine-tine and personalise many interactivity elements in UI/UX designs.

# ACKNOWLEDGEMENTS

Kanij, Madugalla and Grundy are supported by ARC Laureate Fellowship FL190100035.

# REFERENCES

Arora, C., Grundy, J., and Abdelrazek, M. (2023). Advancing requirements engineering through generative ai: Assessing the role of llms. *arXiv preprint arXiv:2310.13976*.

Atlas, S. (2023). Chatgpt for higher education and professional development: A guide to conversational ai.

Benyon, D. (2019). *Designing user experience*. Pearson UK.

Bertão, R. A. and Joo, J. (2021). Artificial intelligence in ux/ui design: a survey on current adoption and [future] practices. *Safe Harbors for Design Research*, pages 1–10.

Bilgram, V. and Laarmann, F. (2023). Accelerating innovation with generative ai: Ai-augmented digital prototyping and innovation methods. *IEEE Engineering Management Review*.

Cooper, A., Reimann, R., and Cronin, D. (2007). *About face 3: the essentials of interaction design*. John Wiley & Sons.

Ekvall, H. and Winnberg, P. (2023). Integrating chatgpt into the ux design process: Ideation and prototyping with llms.

Hosono, S., Hasegawa, M., Hara, T., Shimomura, Y., and Arai, T. (2009). A methodology of persona-centric service design. In *Proceedings of the 19th CIRP Design Conference–Competitive Design*.

Karolita, D., Grundy, J., Kanij, T., Obie, H., and McIntosh, J. (2023). What's in a persona? a preliminary taxonomy from persona use in requirements engineering. In *International Conference on Evaluation of Novel Approaches to Software Engineering 2023*, pages 39–51.

Lewis, J. R. and Sauro, J. (2021). *USABILITY AND USER EXPERIENCE: DESIGN AND EVALUATION*, chapter 38, pages 972–1015. John Wiley & Sons, Ltd.

Liu, P., Yuan, W., Fu, J., Jiang, Z., Hayashi, H., and Neubig, G. (2023). Pre-train, prompt, and predict: A systematic survey of prompting methods in natural language processing. *ACM Computing Surveys*, 55(9):1–35.

Magar, I. and Schwartz, R. (2022). Data contamination: From memorization to exploitation. *arXiv preprint arXiv:2203.08242*.

Main, A. and Grierson, M. (2020). Guru, partner, or pencil sharpener? understanding designers' attitudes towards intelligent creativity support tools. *arXiv preprint arXiv:2007.04848*.

Nguyen-Duc, A., Cabrero-Daniel, B., Przybylek, A., Arora, C., Khanna, D., Herda, T., Rafiq, U., Melegati, J., Guerra, E., Kemell, K.-K., et al. (2023). Generative artificial intelligence for software engineering–a research agenda. *arXiv preprint arXiv:2310.18648*.

Nielsen, L. (2004). Engaging personas and narrative scenarios.

Talebi, S., Tong, E., and Mofrad, M. R. (2023). Beyond the hype: Assessing the performance, trustworthiness, and clinical suitability of gpt3. 5. *arXiv preprint arXiv:2306.15887*.

Tu, N., He, Q., Zhang, T., Zhang, H., Li, Y., Xu, H., and Xiang, Y. (2010). Combine qualitative and quantitative methods to create persona. In *2010 3rd International Conference on Information Management, Innovation Management and Industrial Engineering*, volume 3, pages 597–603.

Ubani, S., Polat, S. O., and Nielsen, R. (2023). Zeroshotdataaug: Generating and augmenting training data with chatgpt. *arXiv preprint arXiv:2304.14334*.

Wang, W., Khalajzadeh, H., Grundy, J., Madugalla, A., McIntosh, J., and Obie, H. O. (2023). Adaptive user interfaces in systems targeting chronic disease: a systematic literature review. *User Modeling and User-Adapted Interaction*, pages 1–68.

White, J., Fu, Q., Hays, S., Sandborn, M., Olea, C., Gilbert, H., Elnashar, A., Spencer-Smith, J., and Schmidt, D. C. (2023a). A prompt pattern catalog to enhance prompt engineering with chatgpt. *arXiv preprint arXiv:2302.11382*.

White, J., Hays, S., Fu, Q., Spencer-Smith, J., and Schmidt, D. C. (2023b). Chatgpt prompt patterns for improving code quality, refactoring, requirements elicitation, and software design. *arXiv preprint arXiv:2303.07839*.

Yu, F., Quartey, L., and Schilder, F. (2023). Exploring the effectiveness of prompt engineering for legal reasoning tasks. In *Findings of the Association for Computational Linguistics: ACL 2023*, pages 13582–13596.

Zhang, X., Liu, L., Wang, Y., Liu, X., Wang, H., Ren, A., and Arora, C. (2023). Personagen: A tool for generating personas from user feedback. In *2023 IEEE 31st International Requirements Engineering Conference (RE)*, pages 353–354.